# Quantum Secure Direct Communication of Continuous-Time Signals Using Whittaker–Nyquist–Shannon Theorem

V. F. Guedes, S. T. de Oliveira, G. L. de Oliveira, J. B. R. Silva, R. V. Ramos

vitorfergue@gmail.com  sergiotahim@gmail.com  glau@ifce.edu.br  joaobrs@ufc.br  rubens.ramos@ufc.br

[1]*Lab. of Quantum Information Technology, Department of Teleinformatic Engineering – Federal University of Ceara - DETI/UFC, C.P. 6007 – Campus do Pici - 60455-970 Fortaleza-Ce, Brazil.*

*Abstract* — **In the present work, we provide a new quantum secure direct communication protocol and its experimental implementation. The proposed protocol can be used to transfer, in a secure way, continuous signals, like audio signal, from Alice to Bob. The security is guaranteed by the quantum nature of optical signals and the Whittaker-Nyquist-Shannon theorem. Furthermore, it can be easily implemented with common optical devices that are commercially available.**

## 1. Introduction

Quantum cryptography is a promising area of data security that guarantees the inviolability of secret data by using the laws of quantum mechanics. Although there exist many quantum cryptographic protocols already proposed like quantum key distribution (QKD) [1-7], quantum secure direct communication (QSDC) [8-15], quantum secret sharing [16-18], quantum digital signatures [19], among others, only few of them have been implemented experimentally. By far, QKD is the most famous and implemented quantum cryptographic protocol and it is already commercially available. The second one is the QSDC protocol, however, there are very few implementations of QSDC protocols, see for example [10,11]. Usually QSDC protocols require single-photons (at least one per bit transmitted), entangled states, quantum memories and only digital information (bits) is transferred from Alice to Bob. Thus, the efficient implementation of QSDC requires good single-photon sources, single-photon detectors, entangled pair sources and quantum memories, depending on the protocol chosen to be implemented. Thus, the massive implementation of QSDC protocols, even in dark optical channels, is still a challenge. In this direction, the present work proposes a new QSDC protocol. Differently of other proposals found in the literature, our protocol transmits in a secure way a continuous signal, like an audio signal, for example, instead of bits, and it uses weak coherent states and homodyne detection, making it completely implementable with low-cost off-the-shelf optical devices.

## 2. Quantum Secure Direct Communication of Continuous Signals

Quantum secure direct communication is an important quantum communication protocol whose goal is to send secret messages in a secure way directly over a quantum channel, that is, without the need of cryptographic keys. Furthermore, QSDC is also a basic cryptographic primitive useful in the construction of other quantum cryptographic protocols, such as quantum dialog and quantum authentication. Despite its importance, the experimental implementation of QSDC protocols found in the literature uses quantum technologies not available or not easily achievable to date in order to permit a massive implementation in real optical networks, even for continuous variable QSDC protocols [20-22]. In this direction, the present work proposes a new QSDC protocol, for secure transmission of continuous signals, that can be easily implementable with current technology, since it uses weak coherent states and homodyne detection. This is possible because its security is based on the on the Whittaker-Nyquist-Shannon theorem.

Basically, the Whittaker–Nyquist–Shannon theorem states that: If a function $x(t)$ contains no frequencies higher than $B$ Hertz, then it can be completely determined from its ordinates at a sequence of points spaced less than $1/(2B)$ seconds apart. Furthermore, the Shannon sampling theory for non-uniform sampling states that a band-limited signal can be perfectly reconstructed from its samples if the average sampling rate satisfies the Nyquist condition. In other words, if a band-limited continuous signal $s(t)$ is randomly sampled with probability $p$, then it can be reconstructed from its samples if the sampling rate is larger than $2f_{max}/p$, where $f_{max}$ is the maximal frequency component in $s(t)$.

The optical setup for running the proposed QSDC protocol is shown in Fig. 1.

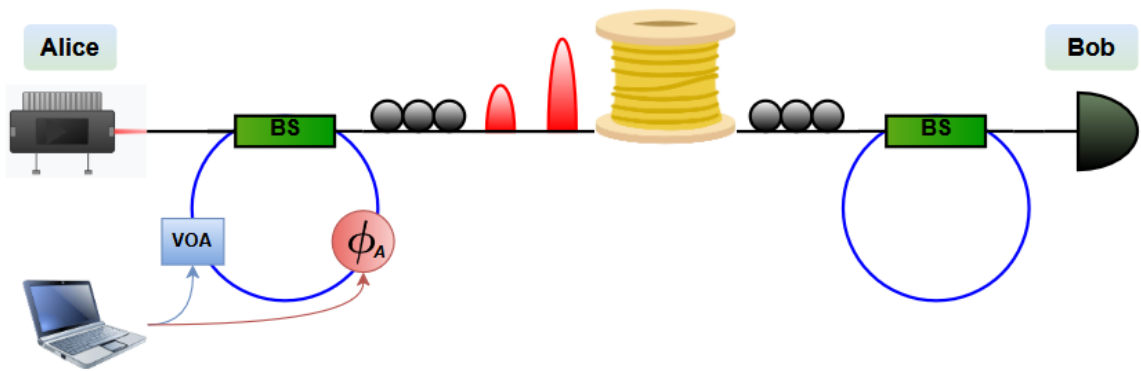

Fig. 1 – Optical setup for quantum secure direct communication of continuous signals. VOA – variable optical attenuator, BS – beam splitter, $\phi_A$ – Alice's phase modulator.

The optical setup shown in Fig. 1 works as follows: Alice sends optical pulses separated by a time interval $T_p$. For each optical pulse sent by Alice the largest part (90%) is transmitted direct towards to Bob, while 10% is reflected into Alice's fiber ring, phase modulated by $\phi_A$ and attenuated by the VOA. This reflected pulse takes a time interval $\tau < T_p$ to make a round trip in Alice's fiber ring. Thus, at the optical channel's input there are two pulses, the first one is a strong pulse that will be used by Bob as local oscillator, followed by a weak coherent pulse having very low mean photon number, $\mu$, and carrying the phase information $\phi$. These two pulses appear again after a time interval $T_p$ and the weaker one can carry new values of $\mu$ and $\phi$. After channel propagation, the first pulse that arrives at Bob (the strong pulse) is partially transmitted to Bob's detector and partially reflected into Bob's fiber ring. This reflected pulse takes a time $\tau$ to travel through Bob's fiber ring and it arrives at Bob's beam splitter at the same moment the weak pulse coming from Alice arrives. They interfere and the result of the interference is observed at Bob's detector. One may note that, differently of common homodyne setups, our scheme uses only one detector and, therefore, the difference between the currents is not realized.

The protocol of QSDC of continuous signals using the setup shown in Fig. 1 is as described below:

1) Alice feeds her phase-modulator with samples of a continuous signal $s(t)$ whose maximal frequency component is $f_{max}$ or samples of another continuous signal $r(t)$ whose maximal frequency component is also $f_{max}$. The signal $r(t)$ works as decoy signal.
2) For each pulse sent by Alice, she chooses randomly, with equal probability, between a sample of $s(t)$, with mean photon number of the weak pulse $\mu_s$, and a sample of $r(t)$ with mean photon number of the weak pulse $\mu_r$.
3) Bob measures the signals sent by Alice using his simple setup shown in Fig. 1.
4) Alice informs to Bob in which time slots she chose the samples of $s(t)$ (or $r(t)$). She also sends to Bob the complete signal $r(t)$.
5) Bob compares the signal $r(t)$ that Alice sent to him with the version of $r(t)$ he reconstructed. If Bob reconstructed correctly the signal $r(t)$, the communication is aborted. If Bob cannot reconstruct the signal $s(t)$, the communication is also aborted.

In order to analyze the security of the proposed protocol, we are going to consider the intercept/resend and beam splitter attacks. Each weak pulse sent by Alice to Bob carries a sample of the signal $s(t)$ or $r(t)$ in its phase. According to Whittaker-Nyquist-Shannon theorem, Bob can reconstruct correctly the signal $s(t)$ only if

$$\frac{1}{T_p} > f_p^s = \frac{4f_{\max}}{1-\exp(-t_c\mu_s)}, \tag{1}$$

where $t_c$ takes into account is the channel's transmissivity that includes also the losses in Bob's apparatus. Thus, hereafter we assume that Alice chooses $T_p$ such that condition (1) is satisfied.

On the other hand, Bob can reconstruct correctly the signal $r(t)$ only if

$$\frac{1}{T_p} > f_p^r = \frac{4f_{\max}}{1-\exp(-t_c\mu_r)}. \tag{2}$$

When Alice chooses $\mu_s > \mu_r$ she can choose $T_p$ such that $s(t)$ is recoverable by Bob but $r(t)$ is not: $f_p^s < T_p^{-1} < f_p^r$. When the eavesdropper Eve intercepts the optical pulses at Alice's output and measures them, she has to choose randomly the mean photon number of the weak pulse she will send to Bob. Choosing between $\mu_s$ and $\mu_r$ with the same probability, the signals $s(t)$ and $r(t)$ will be reconstructed by Bob if

$$\frac{1}{T_p} > f_p^e = \frac{4f_{\max}}{1-\frac{1}{2}\exp(-t_c\mu_s)-\frac{1}{2}\exp(-t_c\mu_r)}. \tag{3}$$

One can easily check that $f_p^s < f_p^e < f_p^r$. If $T_p^{-1}$ is chosen such that $T_p^{-1} < f_p^e$, the intercept/resend attack will make the signal $s(t)$ unrecoverable by Bob, denouncing Eve's presence.

One may note that Eve can change the channel's transmissivity ($t'_c$) in order to compensate her action. In this case, the following relation between the transmissivities (with ($t'_c$) and without ($t_c$) eavesdropping) must be guaranteed by Eve:

$$\frac{4f_{\max}}{1-\frac{1}{2}\exp(-t'_c\mu_s)-\frac{1}{2}\exp(-t'_c\mu_r)} = \frac{4f_{\max}}{1-\exp(-t_c\mu_s)} \Rightarrow \exp(-t'_c\mu_s)+\exp(-t'_c\mu_r) = 2\exp(-t_c\mu_s). \tag{4}$$

Using the Lambert-Tsallis $W_q$ function [23-25] to solve eq. (4), one gets the following new value for the channel's transmissivity that Eve has to guarantee:

$$t'_c = \frac{1}{\mu_s-\mu_r}\ln\left[\frac{\mu_s}{\mu_r-\mu_s}W_{1-\frac{\mu_s}{\mu_r-\mu_s}}\left(\frac{\mu_r-\mu_s}{\mu_s}\left(2e^{-t_c\mu_s}\right)^{\frac{\mu_r-\mu_s}{\mu_s}}\right)\right]. \tag{5}$$

In this case one has $f_p^S = f_p^r < T_p^{-1}$ and, therefore, the signal $r(t)$ will also be reconstructed by Bob, denouncing Eve's presence.

For example, let us consider $\mu_s = 0.6$, $\mu_r = 0.4$ and $t_c = 0.6$. When Eve realizes the intercept/resend attack without changing the channel between Alice and Bob one has: $T_p^{-1} > f_p^S = 3.307 f_{max}$, $f_p^r = 4.68 f_{max}$. Assuming that Alice' chooses $T_p^{-1} = 3.31 f_{max}$, in the absence of Eve's action, Bob can reconstruct $s(t)$ but he cannot reconstruct $r(t)$. On the other hand, if Eve realizes the intercept/resend attack and changes the channel between Alice and Bob in such way that eq. (4) is obeyed, then one has $f_p^r = f_p^S = 3.307 f_{max}$ and Bob will be able to reconstruct both signals $s(t)$ and $r(t)$, what implies that Eve attacked the communication. If Eve realizes the attack but she does not change the channel between Alice and Bob, then $f_p^S = 3.878 f_{max} > T_p^{-1}$ and, therefore, Bob will not be able to reconstruct $s(t)$ what indicates Eve's presence.

For the beam splitter attack Eve, placed close to Alice's output, uses a beam splitter with transmissivity $t_c$ to steal a fraction $(1-t_c)$ of the optical signal sent by Alice. Aiming to compensate the loss she introduced with her beam splitter, Eve also changes the optical channel between Alice and Bob by a lossless optical channel. In this case, according to Whittaker-Nyquist-Shannon theorem, Eve will be able to reconstruct correctly the signal $s(t)$ only if

$$\frac{1}{T_p} > \frac{4 f_{max}}{1 - \exp(-(1-t_c)\mu_s)}. \tag{6}$$

Comparing eqs. (1) and (6) one can note that, knowing the value of $t_c$, Alice can choose a value for $T_p$ that satisfies (1) and does not satisfy (6) only when $t_c > 0.5$. Thus, the proposed QSDC protocol is secure against the beam splitter attack when $t_c > 0.5$, what limits the distance between Alice and Bob to roughly 10 km. For example, once more using $\mu_s = 0.6$, $\mu_r = 0.4$, $t_c = 0.6$ and $T_p^{-1} = 3.31 f_{max}$, one gets $f_p^S = 3.307 f_{max}$ for Bob and $f_p^S = 4.68 f_{max}$ for Eve since her transmissivity is 0.4. Therefore, only Bob will be able to reconstruct the signal $s(t)$.

When the signal to be transmitted by Alice has the required characteristics for sub-Nyquist sampling (signals that are sparse, have limited bandwidth variations or can be effectively compressed), one just changes $2 f_{max}$ in (1)-(6) by $f_{sub}$, the lowest sampling frequency allowed by the sub-Nyquist sampling method used [26-28].

At last, in order to avoid Eve to get values of $s(t)$, Alice separates $s(t)$ in two signals $s_1(t)$ and $s_2(t)$, such that $s(t) = s_1(t) + s_2(t)$. Then she transmits $s_1(t)$ to Bob. If Eve's action is not detected, she transmits $s_2(t)$ to Bob. Having $s_1(t)$ and $s_2(t)$ Bob obtains $s(t)$.

## 4. Experimental Realization of QSDC of Continuous Signals

As a proof of principles, we implemented a simplified version of the optical scheme shown in Fig. 1. The decoy signal $r(t)$ was not used. Figure 2 shows the results of Bob's detection for two different values of phase modulation chosen by Alice (Alice's phase modulator was fed by a train of square pulses). One may note the result of the interference in the amplitude of the weak pulse (on the right side) before attenuation.

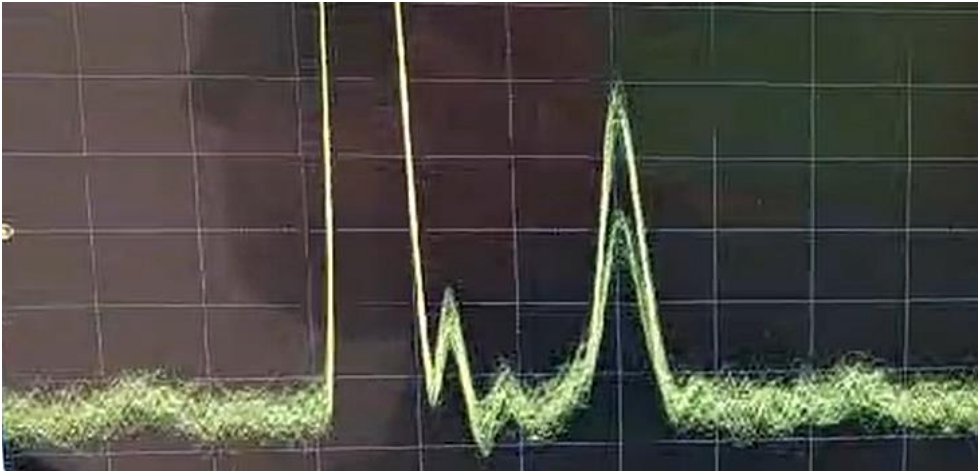

Fig. 2 – Many sequences of two pulses at Bob's output for two different values of $\phi_A$.

Figure 3, by its turn, shows the detection at Bob of a tone of 1.5 MHz (Alice's phase modulator was fed with a 1.5 MHz sinusoidal signal). The mean photon number of the weak pulse used was $\langle n \rangle \approx 0.6$ (1550 nm) at Alice's output, the transmissivity was $t_c \approx 0.6$ and $T_p^{-1} = 5$ MHz. The optical detector used by Bob was PDB465C from Thorlabs.

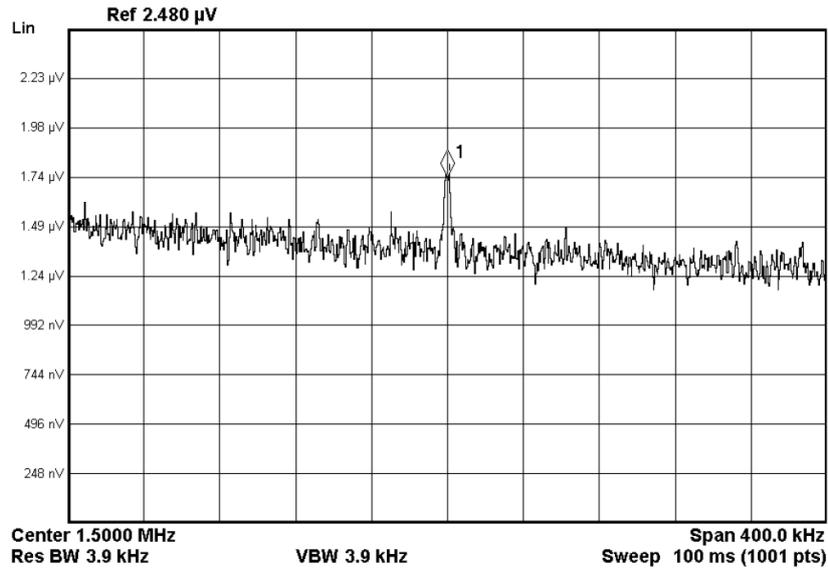

Fig. 3 – Detection of the tone at 1.5 MHz at Bob sent by Alice.

## 5. Conclusions

The QSDC protocol here presented brings two novelties: 1) It permits the secure transmission of continuous signals. 2) Its security is based on the Whittaker–Nyquist–Shannon theorem. Furthermore, it can be easily implemented with common cheap optical devices. Its main disadvantage is the limited channel's length, around 10 km.

## Acknowledgment

This work was partially supported by the Brazilian agencies CNPq (309374/2021-9), CAPES (001) and FUNCAP (ITR-0214-00041.01.00/23). Also, this work was performed as part of the of the project Research, Development and Innovation in Quantum Communication (422316/2023-7).

## Authors contribution

RVR had the main idea and wrote the paper. VFG, STO, GLO and JBRS implemented the experimental QSDC setup.
.

# Data availability

All data generated or analyzed during this study are included in this published article

# Conflict of interest

I declare that the authors have no competing interests as defined by Springer, or other interests that might be perceived to influence the results and/or discussion reported in this paper. On behalf of all authors, the corresponding author states that there is no conflict of interest.

# References

1. K. Inoue, "Differential phase-shift quantum key distribution systems", *IEEE Sel. Top. Quant. Elec.*, vol. 21, no. 3, pp. 6600207, 2015.
2. V. Scarani, H. Bechmann-Pasquinucci, N. J. Cerf, M. Dušek, N. Lütkenhaus, M. Peev, "The security of practical quantum key distribution", *Rev. Mod. Phys.*, no. 81, pp. 1301–1350, 2009.
3. F. Laudenbach, C. Pacher, C.-H. F. Fung, A. Poppe, M. Peev, B. Schrenk, M. Hentschel, P. Walther, H. Hübel, "Continuous-variable quantum key distribution with Gaussian modulation—the theory of practical implementations". *Adv. Quantum Technol.*, no. 1, pp. 1800011/1–37, 2018.
4. Y.-P. Li, W. Chen, F.-X. Wang, Z.-Q. Yin, L. Zhang, H. Liu, S. Wang, D.-Y. He, Z. Zhou, G.-C. Guo, Z.-F. Han, "Experimental realization of a reference-frame independent decoy BB84 quantum key distribution based on Sagnac interferometer", *Opt. Lett*, vol. 44, no. 18, pp. 4523–4526, 2019.
5. G.-Z. Tang, C.-Y. Li, M. Wang, "Polarization discriminated time-bin phase-encoding measurement-device-independent quantum key distribution", *Quantum Engineering*, vol. 3, no. 4, pp. e79, 2021. https://doi.org/10.1002/que2.79
6. X.-F. Wang, X.-J. Sun, Y.-X. Liu, W. Wang, B.-X. Kan, P. Dong, L.-L. Zhao, "Transmission of photonic polarization states from geosynchronous Earth orbit satellite to the ground", *Quantum Eng.*, vol. 3, no. 3, pp. 173, 2021.
7. H.-K. Lo, M. Curty, K. Tamaki, "Secure quantum key distribution", *Nature Photonics*, no. 18, pp. 595–604, 2014.
8. A. G. d. A. H. Guerra, F. F. S. Rios, R. V. Ramos, "Quantum secure direct communication of digital and analog signals using continuum coherent states". *Quantum Inf. Process.*, no. 15, pp. 4747–4758, 2016. https://doi.org/10.1007/s11128-016-1410-0

9. D. Pan, G.-L. Long, L. Yin, Y.-B. Sheng, D. Ruan, S. X. Ng, J. Lu, L. Hanzo, "The evolution of quantum secure direct communication: on the road to Qinternet", *IEEE Communications Surveys & Tutorials*, vol. 26, no. 3, pp. 1898-1948, 2024.
10. J. Y. Hu, B. Yu, M. Y. Jing, *et al.*, "Experimental quantum secure direct communication with single photons". *Light Sci. Appl.*, no. 5, pp. e16144, 2016. https://doi.org/10.1038/lsa.2016.144
11. D. Pan, Z. Lin, J. Wu, H. Zhang, Z. Sun, D. Ruan, L. Yin, G.-L. Long, "Experimental free-space quantum secure direct communication and its security analysis", *Photon. Res.*, no. 8, pp. 1522-1531, 2020.
12. T. Li, G.-L. Long, "Quantum secure direct communication based on single-photon Bell-state measurement", *New J. Phys.*, no. 22, pp. 063017, 2020.
13. X.-J. Li, M. Wang, X.-B. Pan, Y.-R. Zhang, G.-L. Long, "One-photon-interference quantum secure direct communication", *Entropy*, vol. 26, no. 9, pp. 811, 2024. https://doi.org/10.3390/e26090811
14. H. Zhang, Z. Sun, R. Qi, L. Yin, G.-L. Long, J. Lu, "Realization of quantum secure direct communication over 100 km fiber with time-bin and phase quantum states", *Light: Science & Appl.*, vol. 11, no. 83, pp. 1-9, 2022.
15. S. Liu, Z. Lu, P. Wang, *et al.*, "Experimental demonstration of multiparty quantum secret sharing and conference key agreement", *npj Quantum Inf*, no. 9, pp. 92, 2023. https://doi.org/10.1038/s41534-023-00763-z
16. S. Y. Kuo, K. C. Tseng, C. C. Yang, *et al.*, "Efficient multiparty quantum secret sharing based on a novel structure and single qubits", *EPJ Quantum Technol.*, no. 10, pp. 29, 2023. https://doi.org/10.1140/epjqt/s40507-023-00186-x
17. J. Gu, Y.-M. Xie, W.-B. Liu, Y. Fu, H.-L. Yin, Z.-B. Chen, "Secure quantum secret sharing without signal disturbance monitoring", *Opt. Express*, no. 29, pp. 32244-32255, 2024.
18. K. Senthoor, P. K. Sarvepalli, "Theory of communication efficient quantum secret sharing", *IEEE Trans. Inf. Theory*, no. 68, pp. 3164-3186, 2022.
19. R. J. Collins, R. J. Donaldson, V. Dunjko, P. Wallden, P. J. Clarke, E. Andersson, J. Jeffers, G. S. Buller, "Realization of quantum digital signatures without the requirement of quantum memory", *Phys. Rev. Lett.*, vol. 113, no. 4, pp. 040502, 2014.
20. S. Srikara, K. Thapliyal, A. Pathak, "Continuous variable direct secure quantum communication using Gaussian states", *Quantum Inf Process.*, no. 19, pp. 132, 2020. https://doi.org/10.1007/s11128-020-02627-3
21. Z. Cao, Y. Lu, G. Chai, H. Yu, K. Liang, L. Wang, "Realization of quantum secure direct communication with continuous variable, Research, vol. 6, Article ID: 0193, 2023. DOI: 10.34133/research.0193

22. Z. Cao, L. Wang, K. Liang, G. Chai, J. Peng, "Continuous-variable quantum secure direct communication based on gaussian mapping, *Phys. Rev. Appl.*, no. 16, pp. 024012, 2021.
23. G. B. da Silva, R.V. Ramos, "The Lambert-Tsallis Wq function", Phys. A, no. 525, pp. 164-170, 2019.
24. R. V. Ramos, "Analytical solutions of cubic and quintic polynomials in micro and nanoelectronics using the Lambert-Tsallis Wq function", J. Comput. Electron., no. 21, pp. 396-400, 2022. doi.org/10.1007/s10825-022-01852-6
25. J. S. de Andrade, K. Z. Nobrega, R. V. Ramos, "Analytical solution of the current-voltage characteristics of circuits with power-law dependence of the current on the applied voltage using the Lambert-Tsallis Wq function", IEEE Trans. Circuits Syst. II Express Briefs, vol. 69, no. 3, pp. 769-773, 2021. doi.org/10.1109/TCSII.2021.3110407
26. M. Mishali, Y. C. Eldar, "Sub-nyquist sampling: bridging theory and practice", IEEE Signal Process. Mag., no. 11, pp. 98–124, 2011.
27. D. Cohen, Y. C. Eldar, "Spectrum reconstruction from sub-Nyquist sampling of stationary wideband signals", the 10th International Conference on Sampling Theory and Applications (SAMPTA), July 2013.
28. D. Cohen, Y. C. Eldar, "Cyclic spectrum reconstruction from sub-Nyquist samples", the IEEE Globecom 2014, - Signal Processing for Communications Symposium, December 2014.